\documentclass[preprint,preprintnumbers,amsmath,amssymb,nofootinbib]{revtex4}
\usepackage{epsfig}
\usepackage{epstopdf}
\usepackage{bbm}
\usepackage{color}
\newcommand{\be}{\begin{equation}}
\newcommand{\ee}{\end{equation}}
\newcommand{\bea}{\begin{eqnarray}}
\newcommand{\eea}{\end{eqnarray}}
\newcommand{\wtd}{\widetilde}

\def\simlt{\mathrel{\raise.3ex\hbox{$<$\kern-.75em\lower1ex\hbox{$\sim$}}}}
\def\simgt{\mathrel{\raise.3ex\hbox{$>$\kern-.75em\lower1ex\hbox{$\sim$}}}}

\begin{document}
%
\preprint{CUMQ/HEP 197, IPMU18-0136}

\title{A minimal  $U(1)^\prime$ extension of MSSM in light of the B decay anomaly}

\author{Guang Hua Duan$^{1,2}$, Xiang Fan$^{1,2}$, Mariana Frank $^3$, Chengcheng Han$^4$, Jin Min Yang$^{1,2}$ \\~ \vspace*{-0.3cm}}
\affiliation{
$^1$ CAS Key Laboratory of Theoretical Physics, Institute of Theoretical Physics,
 Chinese Academy of Sciences, Beijing 100190, China   \\
$^2$ School of Physical Sciences, University of Chinese Academy of Sciences, Beijing 100049, China\\
 $^3$ Department of Physics, Concordia University, 7141 Sherbrooke St. West, Montreal, QC, Canada H4B 1R6\\
 $^4$ Kavli IPMU (WPI), UTIAS, University of Tokyo,  Kashiwa, Chiba 277-8583, Japan
  \vspace*{1.5cm}
}

\begin{abstract}
Motivated by the $R_K$ and $R_{K^*}$ anomalies from B decays, we extend the minimal supersymmetric model with a
non-universal anomaly-free $U(1)^\prime$ gauge symmetry, coupling
non-universally to the lepton sector as well as
the quark sector. In particular, only the third generation quarks are charged under this $U(1)^\prime$, which can easily
evade the dilepton bound from the LHC searches. An extra singlet is introduced to break this $U(1)^\prime$ symmetry allowing for
the $\mu$-term to be generated dynamically.
The relevant constraints of $B_s-\bar{B}_s$ mixing, $D^0-\bar{D}^0$ mixing and the LHC dilepton searches are considered.
We find that in the allowed parameter space this $U(1)^\prime$ gauge interaction can
accommodate the $R_K$ and $R_{K^*}$ anomalies  and weaken considerably the $Z^\prime$ mass limits  while remaining perturbative up to the Planck scale.

\end{abstract}

\maketitle

\section{Introduction}
\label{sec:introduction}
Supersymmetry (SUSY) remains one of the most compelling frameworks for physics beyond the Standard Model (SM),
which can solve the naturalness problem and provide a natural dark matter candidate.
The simplest supersymmetric model is the Minimal Supersymmetric Standard Model (MSSM)
that contains all the SM particles and their superpartners.
However, the MSSM suffers from the little hierarchy problem (the Higgs boson has a mass of 125 GeV\cite{higgs-cms,higgs-atlas}, indicating significant loop corrections within MSSM), and the $\mu$-problem.  The $\mu$-term parametrizes the coupling between the Higgs bosons at tree-level, and as such, it must be ${\cal O}(1$ TeV) or less. But it is unclear why this is so, since in principle it can be of ${\cal O}(M_{\rm Pl}$) as it does not break SUSY and SM symmetries.

Unfortunately, at present analyses of phenomena at the LHC have yet to yield unambiguous signals of new physics. Perhaps most tantalizing,  there are several interesting excesses in $B$ physics measurements as compared
to SM measurements, of which the theoretically cleanest  is that of
\begin{eqnarray}
&&R_K \equiv \frac{BR(B^\pm \to K^\pm \mu^+ \mu^-)}{BR(B^\pm \to K^\pm e^+e^-)}=0.745^{+0.090}_{-0.074}\pm 0.036 \, , \nonumber\\
&&R_{K^*} \equiv \frac{BR(B^\pm \to K^* \mu^+ \mu^-)}{BR(B^\pm \to K^* e^+e^-)}=0.69^{+0.11}_{-0.07}\pm 0.05\, ,
\end{eqnarray}
 from the LHCb experiment \cite{Aaij:2014ora,Aaij:2017vbb},
which deviate from the SM predictions  by 2.6 $\sigma$ and  2.4-2.5 $\sigma$, respectively. A combined global fits show the deviation from SM could reach $4\sigma$\cite{Altmannshofer:2017yso, DAmico:2017mtc, Capdevila:2017bsm, Hiller:2017bzc, Ciuchini:2017mik, Geng:2017svp}.
These deviations cannot be explained in the MSSM, and may hint that the MSSM symmetry should be extended.

In this work, we consider a non-universal $U(1)^\prime$ gauge extension to the MSSM,  with family-dependent couplings to quarks and leptons.
Such an $U(1)^\prime$ could emerge in GUT, superstring constructions or dynamical electroweak
breaking theories. We adopt a bottom-up approach and take this $U(1)^\prime$ extended supersymmetric
model as a simple extension of MSSM, allowing more flexibility in model parameters.

Note that an important issue for all such $U(1)^\prime$ models is the cancellation of gauge anomalies.
We find an elegant way to achieve this without introducing additional exotics while consistent with the
reported $B$ decay anomalies.
An important feature is that the family-dependence of $U(1)^\prime$ disallows some Yukawa couplings in the superpotential,
leading to massless fermions. The fermion masses, however, can also be induced at loop level via non-holomorphic
operators in the soft sector  as in  \cite{Borzumati:1999sp}.
These non-holomorphic terms can be generated through gravity in mSGURA \cite{Martin:1999hc},
or through gauge mediation in GMSB \cite{ArkaniHamed:1998wc}  via bilinear or trilinear non-holomorphic terms. Since the bilinear non-holomorphic terms \cite{Un:2014afa}
give additional contributions to higgsino mass and may impact the search of natural SUSY at the LHC,
 in this work we consider the trilinear non-holomorphic
terms \cite{Chattopadhyay:2017qvh,Chattopadhyay:2016ivr}.
Furthermore,  an extra singlet chiral field is needed to break the $U(1)^\prime$ symmetry, which can simultaneously
generate the $\mu$-term and hence solve the $\mu$-problem.
Finally, due to the  contributions of the additional scalar  to the Higgs mass, the fine-tuning is also alleviated in this model.

This work is organized as following. In Sec.~\ref{sec:model} we introduce a non-universal and gauge anomaly-free $U(1)^\prime$
 extention of the MSSM. 
In Sec.~\ref{sec:anomalies} we investigate the allowed parameter space to explain the $B$ decay anomalies
and discuss the relevant experimental constraints on this model. Finally, we draw our conclusions in Sec.~\ref{sec:conclusion}.

\section{A non-universal $U(1)^\prime$ extension of MSSM}
\label{sec:model}

The process of $B$ decay can be described by the effective Hamiltonian:
\begin{eqnarray}
{\cal H}_{eff} &=&-\frac{4G_F}{\sqrt{2}}V_{tb}V_{ts}^*[C_9{\cal O}_9+C_{10}{\cal O}_{10}]+h.c
\end{eqnarray}
with
\begin{eqnarray}
{\cal O}_9 &=&\frac{e^2}{16\pi^2}\bar{s}_L\gamma^\mu b_L\bar{\ell}\gamma_\mu\ell,
~~~~{\cal O}_{10} =\frac{e^2}{16\pi^2}\bar{s}_L\gamma^\mu b_L\bar{\ell}\gamma_\mu\gamma^5\ell \, .
\end{eqnarray}
The SM predicts $C_9 \sim -C_{10} \sim 4.1$. To accommodate the $B$ decay anomaly,  contributions from new physics are required.
From the studies in \cite{Altmannshofer:2017yso}, a Wilson coefficient $\delta C_9^\mu\in[-2.12,-1.1] ([-2.87,-0.7])$
(assuming $\delta C^\mu_{10}=0$) is favored in 1(2) $\sigma$ region. This indicates that the new particles should
couple to the left-handed down type quarks.

In this work, we interpret the new physics contribution  from an additional $U(1)^\prime$ gauge symmetry extension of
the MSSM.  Besides the chiral multiplets in the MSSM,the $U(1)^\prime$ extension also introduce chiral multiplets of three
right-handed neutrinos $\nu^c_{1,2,3}$, as required by the neutrino mass, and a new singlet $S$ which
breaks the $U(1)^\prime$ gauge symmetry.

First, to evade the bounds on the $Z^\prime$ mass from LHC dilepton search, one possibility is that the $U(1)^\prime$ would have very weak couplings
with the first two generations of quarks. As the minimal set up, we require that the $U(1)^\prime$ only couple
to the third generation quarks. Even so, there are a lot of possibilities for such an anomaly-free $U(1)$ extension,
such as $(B-L)_3$ \cite{Alonso:2017bff, Alonso:2017uky} and $L_i-L_j$ $(i,j=1,2,3)$ \cite{Altmannshofer:2014cfa,Duan:2017qwj,Tang:2017gkz, Yin:2018qcs}.

Second, the $U(1)^\prime$ should
couple to both lepton and quark sectors, which excludes the possibility of $L_i-L_j$ models.
Interestingly, $a(B-L)_3+b(L_i-L_j)$ is also anomaly free and is sufficient to satisfy the previous requirements.
In particular, we focus on $a(B-L)_3+b(L_\mu-L_\tau)$\footnote{We also assume the right-handed neutrino takes the same
charge as the corresponding lepton, which is a little different from the original $L_\mu-L_\tau$ symmetry.},
where the muon coupling is present without affecting the electron sector.  For the Yukawa structure
of the models, we will show that  the combination of $(a, b)$  is unique, up to a global factor.

As required by the explanation of $B$ decay anomalies,  the couplings of $b_L$ and $s_L$ corresponding to the $U(1)^\prime$ are needed.
This can be realized by rotating the flavor eigenstates to the mass eigenstates. Without triggering large mixing in right-handed down quark sector, such a rotation requires a non-zero
$m_d^{23}$ in the mass matrix of down type quarks.   This mass term can arise from the superpotential
 \begin{equation}
W\supset Y_d^{23} Q_2H_d D_3^c
\end{equation}
which means that
\begin{eqnarray}
\label{d3cq2hd}
Q_{D_3^c}+ Q_{Q_2}+Q_{H_d} =0 ~~\Rightarrow~~ Q_{H_d}=-Q_{D_3^c}=Q_{Q_3^c}=a/3 \, .
\end{eqnarray}
The charge of $Q_{H_d}$ seems to contradict  the assumption of the model $a(B-L)_3+b(L_\mu-L_\tau)$.
Noticing that $H_d$ has the same SM quantum charges as $L_3$, we can exchange $H_d \leftrightarrow L_3$
without causing additional anomaly problems if  $a/3= -a-b$ is satisfied,
and thus we derive the unique solution $b=-3/4 a$.
At the same time, we can also exchange $S \leftrightarrow \nu^c_3$. Then the Dirac mass term of
neutrinos in the superpotential becomes
\begin{equation}
W\supset \lambda_s SH_uH_d \, .
\end{equation}
With the spontaneous symmetry breaking of   $U(1)^\prime$ ($S$ developing a non-zero VEV ),
not only the $U(1)^\prime$ is broken,  but also the  $\mu$-term is generated dynamically.

In the following, we take $a=3/2$ and summarize $U(1)^\prime$ charge of matter chiral superfields
in Table \ref{zp:tab:qn}. After requiring $R$-parity conservation, the  most general superpotential
is given by
\begin{eqnarray}
W&=& Y_u^{ij} Q_iH_uU^c_j+Y_u^{33}Q_3H_uU^c_3 -Y_d^{13}Q_1H_d D^c_3-Y_d^{23}Q_2H_d D^c_3 +Y_{\nu}^{11} L_1H_u\nu^c_1 + Y_{\nu}^{13} L_1H_u\nu^c_3 \nonumber \\
 &&+ Y_{\nu}^{31} L_3H_u\nu^c_1 + Y_{\nu}^{33} L_3H_u\nu^c_3+ Y_{\nu}^{22} L_2
H_u\nu^c_2-Y_e^{33}L_3H_d E^c_3 + M \nu^c_{1,3} \nu^c_{1,3}  + \lambda_s S H_u H_d
 \label{zp:eq:supo2}
\end{eqnarray}
where $i,j=1,2$ and  the last term induces an effective $\mu$ parameter $\lambda_s v_s/\sqrt{2}$ when the singlet Higgs $S$
acquires a VEV, $\langle S \rangle=v_s /\sqrt{2}\sim {\cal{O}}({\rm TeV})$, providing a dynamical solution to the $\mu$ problem.

\begin{table}[h]
\caption{
Quantum numbers for $a=3/2$ under the  anomaly-free $U(1)^\prime$ gauge group in the model. }
\label{zp:tab:qn}
\begin{center}
\begin{tabular}{|c|c||c|c||c|c|} \hline
$Q_{1,2}$   & 0        & $L_1$ ,$E_1^c$       &  0    & $S$ &-1/2\\ \hline
$Q_3$       & +1/2       &  $L_2$       &  -2   & $\nu_{2}^c$  & +2\\ \hline
$U_{1,2}^c$ ,$D_{1,2}^c$  & 0        &  $L_3$       &  0   & $\nu_{1}^c,\nu_{3}^c$   &0\\ \hline
$U_3^c$       & -1/2       &  $E_{2}^c$   &  +2   & $H_u$        &0\\ \hline
$D_3^c$& -1/2       &  $E_3^c$       & -1/2    & $H_d$        & +1/2   \\ \hline
\end{tabular}
\end{center}
\end{table}

The family dependence of the $U(1)^\prime$ invariance necessarily forbids certain Yukawa couplings in
the superpotential, rendering to some  fermions massless. The requisite fermion masses, however, can be
induced at loop level via non-holomorphic operators in the soft sector.
This means that, in addition to the above terms,  the Lagrangian must contain some non-holomorphic
SUSY breaking terms:
 \begin{eqnarray}
 \label{mssmnonholo}
   -{\mathcal{L}}^{\rm non-holomorphic}_{soft}&=&
C_E^{11}H_{u}^* \wtd l_1 \wtd E_{R1}^{c}+C_E^{31}H_{u}^* \wtd l_3 \wtd E_{R1}^{c}+C_E^{22}H_{u}^* \wtd l_2 \wtd E_{R2}^{c}+
 C_U^{31} H_{d}^* \wtd q_3\wtd u_R^{c} \nonumber  \\
&& +C_U^{32} H_{d}^* \wtd q_3\wtd c_R^{c}
   +C_D^{11} H_{u}^* \wtd q_1\wtd d_R^{c} +C_D^{12} H_{u}^* \wtd q_1\wtd s_R^{c}+C_D^{21} H_{u}^* \wtd q_2\wtd d_R^{c}
\nonumber \\
&& +C_D^{22} H_{u}^* \wtd q_2\wtd s_R^{c}+C_D^{33} H_{u}^* \wtd q_3\wtd b_R^{c}+h.c.
\end{eqnarray}

\begin{figure}[h]
  \centering
  \includegraphics[width=4in]{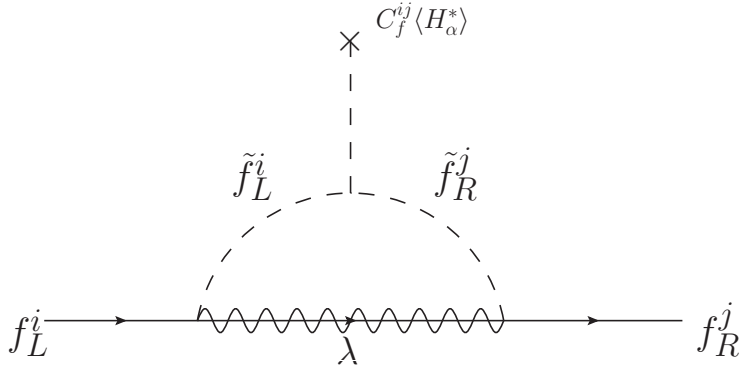}\\
   \caption{Feynman diagram of fermion mass generation from non-holomorphic terms, where $H_\alpha=H_u,H_d$.  }\label{feyn}
   \end{figure}
Now the down-type quarks, for instance, can obtain finite masses via triangular diagrams shown in Fig.\ref{feyn}
with  $\wtd D_L$, $\wtd D_R^c$ and a neutral gaugino $\lambda$ in the loops, whose magnitude is proportional to $C_D$.
This radiative mechanism for generating fermion masses is generic, with the coupling to the `wrong' Higgs doublet
in (\ref{mssmnonholo}) being essential for giving mass to fermions. The origin of non-holomorphic `soft' terms
can be induced from supergravity mediated supersymmetry breaking \cite{Martin:1999hc}. Therefore, the graphs in Fig.\ref{feyn}
induce fermion mass matrix elements at one-loop level. For $m_f\ll m_{\tilde{f}},m_\lambda$, the one-loop contribution
to fermion mass is given by \cite{Borzumati:1999sp}
 \begin{eqnarray}
m_f&=&C_f v_\alpha[\frac{\alpha_s}{2\pi}\xi_fm_{\tilde{g}}
I(m_{\tilde{f}_1},m_{\tilde{f}_2},m_{\tilde{g}})+\frac{\alpha_Y}{2\pi}
\sum_{j=1}^6 K_f^jm_{\chi_j^0}I(m_{\tilde{f}_1},m_{\tilde{f}_2},m_{\chi_j^0})] ,
 \end{eqnarray}
where $\alpha_Y=g_1^2/4\pi$, $v_a=v_u(v_d)$ for down(up) type fermions and
$\xi_f = 4/3,0$ for quarks and leptons, respectively.
The  loop function $I(m_{\tilde{f}_1},m_{\tilde{f}_2},m_{\chi_j^0})$ and
the coupling coefficients $K^j_f$ are given by
\begin{eqnarray}
&&I(m_{\tilde{f}_1},m_{\tilde{f}_2},m_{\chi_j^0})=\frac{1}{m_{\tilde{f}_1}^2-m_{\tilde{f}_2}^2}[\frac{\ln (m_\lambda^2/ m_{\tilde{f}_1}^2)}{m_\lambda^2/ m_{\tilde{f}_1}^2-1}-\frac{\ln (m_\lambda^2/ m_{\tilde{f}_2}^2)}{m_\lambda^2/ m_{\tilde{f}_2}^2-1}] ,\\
&&K_f^j=[Y_{fR}N_{jB}+\frac{g_E}{g_1}Q_{f,R}N_{j,Z^\prime}][Y_{fL}N_{jB}+\frac{g_E}{g_1}Q_{f,L}N_{j,Z^\prime}+\cot\theta N_{jW}T_{f,L}^3]
\end{eqnarray}
with $Y_f$ being the $U(1)$ hypercharge, $g_E$ and $Q_f$ being the $U(1)^\prime$  gauge coupling and charge for fermion $f$.
The $\tilde{Z}^\prime$,  bino and wino components of the $j$-th neutralino are expressed in terms of the diagonalizing neutralino mass matrix $N$ as $N_{jB^\prime}, N_{jB}$ and $N_{jW}$ ,
respectively. We denote the loop induced fermion ($f$) mass matrix elements with $\kappa _f^{ij}v_\alpha$.

Typically, we can take $C_D^{33}=4$ TeV, $C_E^{22}=-4$ TeV,  $m_{\chi_j^0}=2$ TeV, $\tan\beta=10,\, g_E=0.2$ and the sbottom , smuon
and gluino mass $m_{\tilde{b}_{1,2}, \tilde{\mu}_{1,2}}=m_{\tilde{g}}=3$ TeV. Then, one can get the fermion mass matrix elements
$\kappa_d^{33}v_u=4.06$ GeV and $\kappa_e^{22}v_u=0.1$ GeV. Unless the non-holomorphic terms are  large, the top quark
and tau lepton mass cannot be easily obtained. Fortunately, top quark and tau lepton masses have been generated already
at tree level via Yukawa coupling terms. Therefore, we can set the proper Yukawa couplings and non-holomorphic terms
such that all fermions get correct masses. For example, the mass matrices of the up and down quark sectors are written as
 \begin{eqnarray}
m_u =\left(
   \begin{array}{cc}
    Y_u^{ij}v_u/\sqrt{2} &0\\
    m_u^{NH}&Y_u^{33}v_u/\sqrt{2} \\
   \end{array}
 \right),~
 m_d =\left(
   \begin{array}{cc}
    \kappa_d^{ij}v_u &(m_d^{Y})^T\\
    0&\kappa_d^{33}v_u \\
   \end{array}
 \right)~
 \end{eqnarray}
 where $i,j=1,2$, $m_d^{Y}=(Y_d^{13},Y_d^{23})v_d/\sqrt{2}$, and $m_u^{NH}=(\kappa^{31}_uv_d,\kappa^{32}_uv_d)$ are obtained
from the non-holomorphic one-loop corrections.
Note that there is a texture structure in the mass matrix, and the 0-terms cannot be derived from either the Yukawa terms
or from the non-holomorphic terms.
Similarly, one can also write down the mass matrix in the lepton sector.
After diagonalizing the quark mass matrices,  we  derive the correct quark masses
  \begin{eqnarray}
  V_{u,L}^\dagger m_u V_{u,R}&=&diag\{m_u,m_c,m_t\} \\
  V_{d,L}^\dagger m_d V_{d,R}&=&diag\{m_d,m_s,m_b\}
  \end{eqnarray}
where $ V_{u(d),L(R)}$ are unitary rotation matrices. To accommodate the $B$ meson decay anomaly in our model,
a small mixing between the second and third generations is needed in $V_{d,L}$.
Specifically, we require $V_{d,L}=R_{23}(\theta_q)$, where
\begin{eqnarray}
R_{23}(\theta_q) =\left(
   \begin{array}{ccc}
    1 &0&0\\
    0&c_{\theta_q}&s_{\theta_q}\\
    0&-s_{\theta_q} &c_{\theta_q}\\
   \end{array}
 \right)~.
 \end{eqnarray}
Then we can obtain $V_{u,L}=R_{23}(\theta_q)V_{\rm CKM}^\dagger$ such that $V_{u,L}^\dagger V_{d,L}=V_{\rm CKM}$. And $V_{u(d),R}$
 can also be fixed at the same time.

Since $H_d$ is charged under the $U(1)^\prime$ group, after symmetry breaking it will induce $Z-Z^\prime$ mixing at tree level.
The mass matrix of gauge boson under gauge eigenstates $(A_\mu^Y,W_\mu^3,A_\mu^\prime)$ is given by
 \begin{eqnarray}
 {\cal M} ^2=\left(
   \begin{array}{ccc}
    \frac{g_1^2v^2}{4} &-\frac{1}{4}g_1g_2v^2&-\frac{1}{4}g_1g_E v^2\cos^ 2\beta\\
    -\frac{1}{4}g_1g_2v^2& \frac{g_2^2v^2}{4}&\frac{1}{4}g_2g_Ev^2\cos^2\beta\\
    -\frac{1}{4}g_1g_E v^2\cos^ 2\beta&\frac{1}{4}g_2g_Ev^2\cos^2\beta &\frac{1}{4}g_E^2(v_s^2+v^2\cos^2\beta)\\
   \end{array}
 \right)\\\nonumber
 \end{eqnarray}
 where $g_E$ is the coupling of the $U(1)^\prime$ gauge group, and we neglect $Z-Z^\prime $ kinetic mixing \cite{Hook:2010tw}. The mass matrix can be diagonalized by matrix ${\cal V}$
to obtain mass eigenstates $(\gamma, Z, Z^\prime)$ :
 \begin{eqnarray}
 {\cal V} =\left(
   \begin{array}{ccc}
    c_\theta &-s_\theta c_{\theta^\prime}&s_\theta s_{\theta^\prime}\\
    s_\theta& c_\theta c_{\theta^\prime}&-c_\theta s_{\theta^\prime}\\
    0&s_{\theta^\prime} &c_{\theta^\prime}\\
   \end{array}
 \right)\\\nonumber
 \end{eqnarray}
 with
 \begin{eqnarray}
&& {\cal V}^T{\cal M} ^2{\cal V} = {\rm diag} \{0,m_Z^2,m_{Z^\prime}^2\},\\
&&\tan\theta = \frac{g_1}{g_2}~,\\
&&\tan 2\theta^\prime=-\frac{2M^2_{Z-Z^\prime}}{\frac{1}{4}[g_E^2(v_s^2+v^2\cos^2\beta)-(g_1^2+g_2^2)v^2]},
\end{eqnarray}
where $\theta$ is the SM weak angle and $Z-Z^\prime$ mixing
$M^2_{Z-Z^\prime}= \frac{1}{2}\sqrt{g_1^2+g_2^2}v^2\cos^ 2\beta$.
The $Z$ boson mass measurement requires  \cite{pdg}
\begin{eqnarray}
|\frac{m_Z^2-m_{Z^0}^2}{m_{Z^0}^2}|=|-t_{\theta^\prime}^2\frac{m_{Z^\prime}^2-m_{Z^0}^2}{m_{Z^0}^2}|<4.6\times10^{-5}
\end{eqnarray}
where $m_{Z^0}^2=c_{\theta^\prime}^2m_Z^2+s_{\theta^\prime}^2m_{Z^\prime}^2 $ with  $m_{Z^0}^2=\frac{1}{4}(g_1^2+g_2^2)v^2$
at tree level.  We find for a moderate $\tan\beta$ and TeV scale $Z^\prime$ mass, $\theta^\prime$ is
sufficiently small, so we ignore the mixing term in the following discussion.
Then the fermions couple to $Z^\prime$ through the current $J_{Z^\prime}^\mu$:
\begin{eqnarray}
-{\cal L}&\supset&J_{Z^\prime}^\mu Z^\prime_\mu
\end{eqnarray}
where
\begin{eqnarray}
J_{Z^\prime}^\mu&=&g_E\sum_f [Q_f \bar{\psi}_{f}\gamma^\mu\psi_{f}]
\end{eqnarray}
with  $Q_f$ denoting  the fermion  $U(1)^\prime$  charge.

\section{Resolving  $B$ decay anomalies }
\label{sec:anomalies}
\begin{figure}[ht]
  \centering
  \includegraphics[width=3.3in]{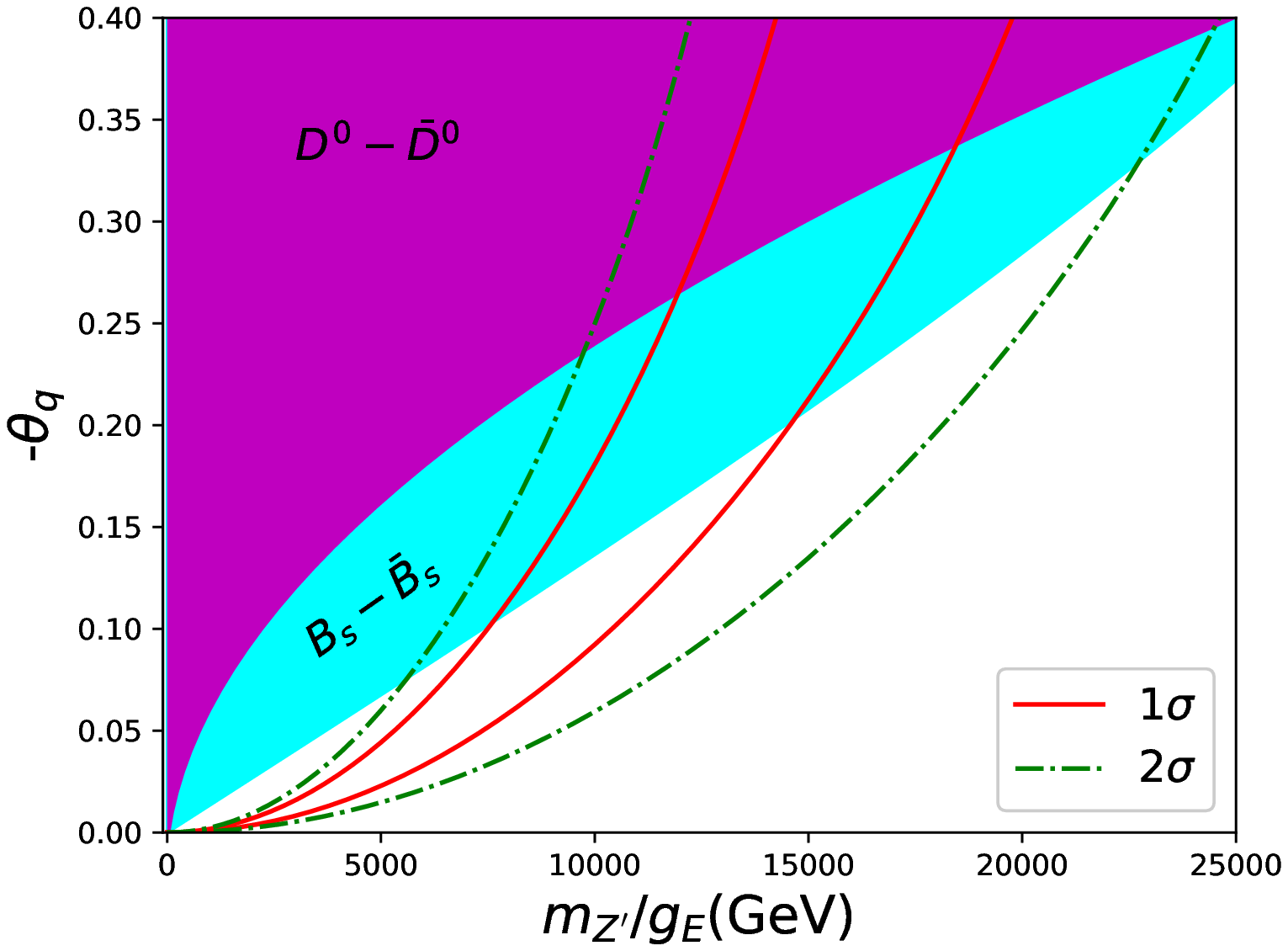}
  \includegraphics[width=3.3in]{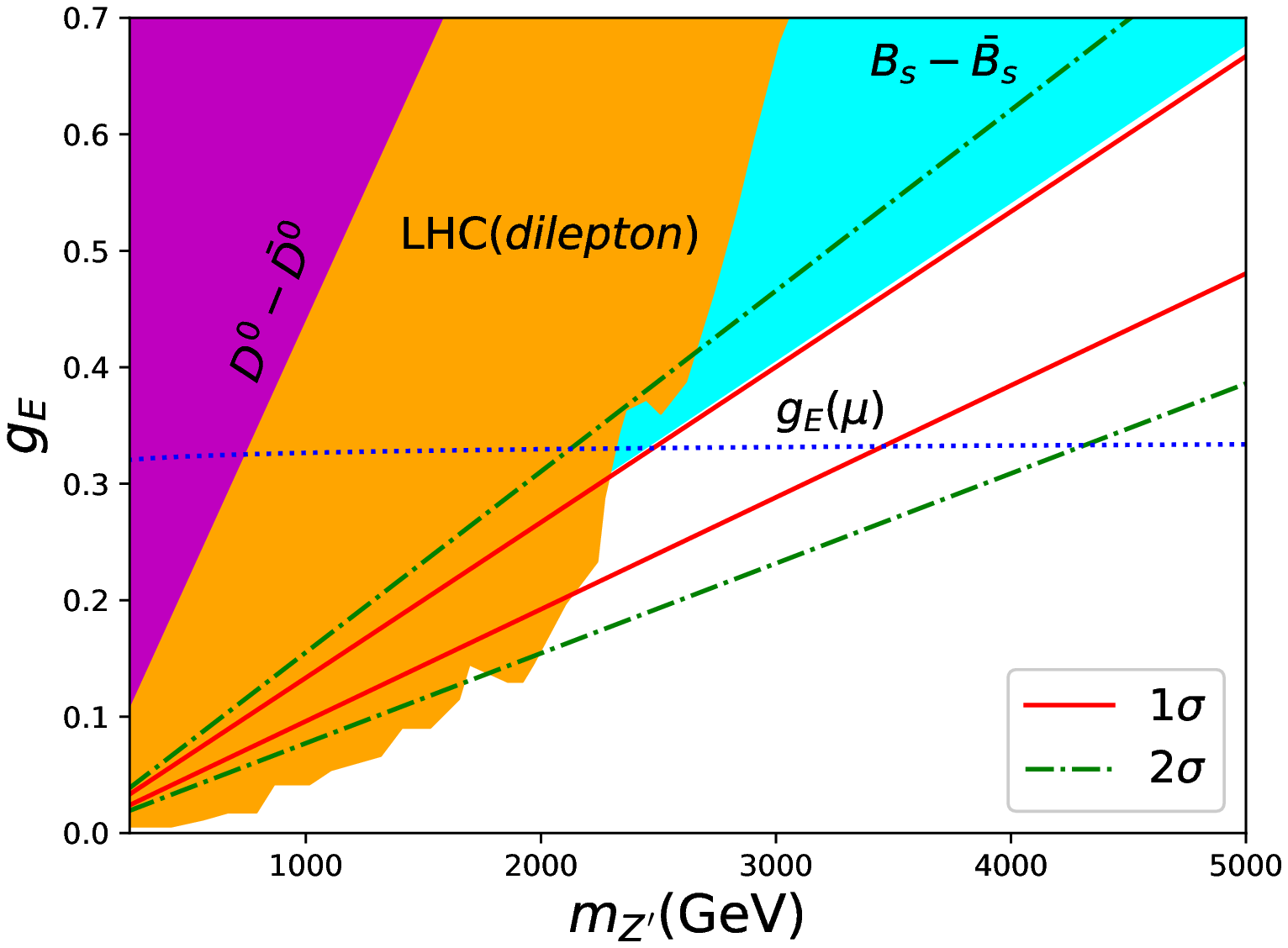}\\
\vspace{-0.5cm}
\caption{
The region between the two solid (dashed) curves
can explain $R_K,R_{K^*}$ anomaly at 1(2)$\sigma$ level. The shaded regions are excluded
by $B_s-\bar{B}_s$ mixing, $D^0-\bar{D}^0$ mixing and the LHC dilepton searches  \cite{Aaboud:2017buh} for $\theta_q=-0.1$.
The region above the dotted horizontal curve in the right panel
is excluded by the Landau pole requirement.}\label{banomaly}
\end{figure}

The coefficients of $C_{10}^\mu$ vanish due to the vector-like coupling of $Z^\prime$ to muon in this model. The contribution of ${\cal O}_9$ to $B$ meson decay can be obtained by integrating out  $Z^\prime$. The effective Lagrangian of ${\cal O}_9$ is given by
\begin{eqnarray}
{\cal L}_{{\cal O}_9} &=&-\frac{c_{\theta_q}s_{\theta_q}}{m_{Z^\prime}^2}g_E^2\bar{s}_L\gamma^\mu b_L\bar{\mu}\gamma_\mu\mu
\end{eqnarray}
yielding
\begin{eqnarray}
C_9^\mu &=&-\frac{\pi}{\alpha\sqrt{2}G_FV_{tb}V_{ts}^*}\frac{2c_{\theta_q}s_{\theta_q}}{m_{Z^\prime}^2}g_E^2 \, .
\end{eqnarray}
The allowed range of the Wilson coefficient is $C_9^\mu\in[-2.12,-1.1]([-2.87,-0.7])$
at 1(2) $\sigma$ level \cite{Altmannshofer:2017yso}. We show the best-fit range in the left-hand panel of Fig.~\ref{banomaly}.
The non-universal couplings in the down quark sector also contribute to $B_s-\bar{B}_s$ and $D^0-\bar{D}^0$ mixings.
The relevant effective operators for these mixings  are given by
\begin{eqnarray}
{\cal L} &=&-\frac{c^2_{\theta_q}s^2_{\theta_q}}{8m_{Z^\prime}^2}g_E^2(\bar{s}_L\gamma^\mu b_L)^2
-\frac{g_E^2c_D^2}{8m_{Z^\prime}^2}(\bar{u}_L\gamma^\mu c_L)^2
\end{eqnarray}
with
\begin{eqnarray}
c_{D} &=&(c_{\theta_q}V_{ub}-s_{\theta_q}V_{us})(c_{\theta_q}V_{cb}^*-s_{\theta_q}V_{cs}^*)\, .
\end{eqnarray}
The global fit of $B_s-\bar{B}_s$ mass mixing together with the CKM fit
give an upper bound $ |g_E^2 c^2_{\theta_q}s^2_{\theta_q}/8m_{Z^\prime}^2|<1/(210{\rm TeV})^2$ \cite{Arnan:2016cpy}.
In Fig.~\ref{banomaly} one can see that the $B_s-\bar{B}_s$ mixing measurement requires
the mixing angle $|\theta_q|<0.3$ for $m_Z^\prime/g_E=20$ TeV.
In the mean time, the constraints from $D^0-\bar{D}^0$ mixing require $|g_E^2 c^2_D /8m_{Z^\prime}^2|<1/(1900{\rm TeV})^2$ \cite{Arnan:2016cpy} and thus
it can be seen that the $D^0-\bar{D}^0$ mixing constraints are weaker than the $B_s-\bar{B}_s$ mixing
due to the additional CKM  suppression of $c_D$.

Since $Z^\prime$  couples only to the third generation quarks directly, its production rate is suppressed by
parton distribution functions at the LHC. However, the branching ratio $ Br(Z^\prime\rightarrow\mu^+\mu^-)$
is large, so that $\sigma(pp\rightarrow Z^\prime)\times Br(Z^\prime\rightarrow\mu^+\mu^-)$
is sizable. We use \textsf{MadGraph5\_aMC@NLO} \cite{Madgraph} to calculate the production cross section.
It can been seen from  Fig.\ref{banomaly} that the LHC dilepton searches can cover $m_{Z^\prime}<2.5$ TeV for $g_E\sim 0.3$. This is a sizable reduction in the expected mass of the $Z^\prime$ from the ATLAS expectation, assuming $U(1)^\prime$ gauge extensions with universal couplings \cite{Aaboud:2017buh}, where $M_{Z^\prime} \simlt 4.1$ TeV.

 We explore next the allowed values for the $U(1)^\prime$ gauge coupling $g_E$. If the coupling $g_E$ is too large at the renormalization scale, it will encounter the Landau pole and blow up
at some higher energy scale. The one loop beta function of $U(1)^\prime$ gauge coupling is given by
\begin{eqnarray}
\beta(g_E)&=&\frac{g_E^3}{16\pi^2}\sum_iQ_i^2 =\frac{5g_E^3}{4\pi^2} \, .
\end{eqnarray}
If the Landau pole occurs at scale $\Lambda$, expected
to be $\Lambda=1.2\times10^{19}$GeV (Planck Mass),
then
\begin{eqnarray}
g_E(\mu)&=&(\frac{5}{2\pi^2}\log \frac{\Lambda}{\mu})^{-1/2} \sim 0.3
\end{eqnarray}
with $\mu=m_Z^\prime$. This means that the $U(1)^\prime$ gauge coupling remains perturbative up to the
Planck mass scale if $g_E<g_E(\mu)$.

Before ending this section, we would like to make some comments:
\begin{itemize}
\item[(i)]  The F-term of the additional singlet boson $S$ and the D-term of $H_d$ contribute to Higgs boson mass at tree level\cite{Demir:2005ti}.
In the decoupling limit, $(m_h^2)_{\rm tree}=m^2_Z\cos^2\beta+\frac{1}{2}\lambda_s^2v_s^2\sin^22\beta+\frac{1}{4}g_E^2(v_s\cos^2\beta )^2$, so that the fine-tuning in this model can be alleviated, as stated in Section \ref{sec:introduction}.
\item[(ii)] The $Z^\prime$ contribution to $(g-2)_\mu$ is very small. Despite the fact that $Z^\prime$ couples strongly to muons, its mass is still relatively heavy, $M_{Z^\prime} \sim 2.5$ TeV.  Nonetheless, $(g-2)_\mu$ can receive sizable
contributions from the supersymmetric sector of the model, especially smuon or chargino loop corrections as in the MSSM \cite{Cox:2018qyi}.
\item[(ii)] The extra singlino and gaugino may affect the components of the LSP (DM candidate). However,
the VEV of $S$ provides a mass term $\sim M_{Z^\prime} $ around TeV scale, and therefore the dark matter sector of this $U(1)^\prime$ with non-universal couplings is not essentially changed from  the MSSM.
\end{itemize}

\section{Conclusions}
\label{sec:conclusion}

Motivated by the $R_K$ and $R_{K^*}$ anomalies from B decays, we extended the MSSM with a non-universal anomaly-free
$U(1)^\prime$ gauge symmetry.   The model is rendered anomaly free without introduction of exotics. Up-type quarks acquire masses as usual via Yukawa couplings, while down-type quarks require the presence of non-holomorphic terms at loop-level.
An extra singlet is introduced to break this $U(1)^\prime$ symmetry, whose VEV can simultaneously
generate the $\mu$-term. We showed that a $Z^\prime$ neutral gauge boson from this $U(1)^\prime$ model with a mass 3-4 TeV can accommodate
the $R_K$ and $R_{K^*}$ anomalies while remaining perturbative up to the Planck scale.
The relevant constraints from $B_s-\bar{B}_s$ and $D^0-\bar{D}^0$
mixings as well as the LHC dilepton searches were also considered.  The model presented is very predictive, restricting relevant variables in a small range of parameter space. The allowed mixing between the second and third generations is  $|\theta_q |\simlt 0.3$ . Perturbativity to Planck scale requires the $U(1)^\prime$ coupling constant to be $g_E \simlt 0.3$, while the $Z^\prime$ mass is lowered to $\simgt 2.5$ TeV for $\theta_q = -0.1$.

\section*{Acknowledgment}
This work is supported by the National Natural Science Foundation of China (NNSFC) under
grant No. 11675242,
by Peng-Huan-Wu Theoretical Physics Innovation Center (11747601),
by the CAS Center for Excellence in Particle Physics (CCEPP),
by the CAS Key Research Program of Frontier Sciences
and by a Key R\&D Program of Ministry of Science and Technology under number 2017YFA0402200-04. MF acknowledges the NSERC for partial financial support under grant number SAP105354, and thanks CAS Key Laboratory of Theoretical Physics, Institute of Theoretical Physics,
 Chinese Academy of Sciences, for hospitality, while part of this work was completed.


\begin{thebibliography}{99}
 \bibitem{higgs-atlas}
G. Aad {\it et al.}\ (ATLAS Collaboration), Phys. Lett. B \textbf{710}, 49
(2012).

\bibitem{higgs-cms}
S. Chatrachyan {\it et al.}\ (CMS Collaboration), Phys. Lett. B
\textbf{710}, 26 (2012).

 \bibitem{Aaij:2014ora}
  R.~Aaij {\it et al.} [LHCb Collaboration],
  Phys.\ Rev.\ Lett.\  {\bf 113}, 151601 (2014)
  [arXiv:1406.6482 [hep-ex]].
  \bibitem{Aaij:2017vbb}
  R.~Aaij {\it et al.} [LHCb Collaboration],
  JHEP {\bf 1708}, 055 (2017)
  [arXiv:1705.05802 [hep-ex]].



\bibitem{Altmannshofer:2017yso}
  W.~Altmannshofer, P.~Stangl and D.~M.~Straub,
  Phys.\ Rev.\ D {\bf 96} (2017) no.5,  055008
  [arXiv:1704.05435 [hep-ph]].
  
\bibitem{DAmico:2017mtc}
  G.~D'Amico, M.~Nardecchia, P.~Panci, F.~Sannino, A.~Strumia, R.~Torre and A.~Urbano,
  JHEP {\bf 1709} (2017) 010
  [arXiv:1704.05438 [hep-ph]].
  
 \bibitem{Capdevila:2017bsm}
  B.~Capdevila, A.~Crivellin, S.~Descotes-Genon, J.~Matias and J.~Virto,
  JHEP {\bf 1801} (2018) 093
  [arXiv:1704.05340 [hep-ph]].
  
\bibitem{Hiller:2017bzc}
  G.~Hiller and I.~Nisandzic,
  Phys.\ Rev.\ D {\bf 96} (2017) no.3,  035003
  [arXiv:1704.05444 [hep-ph]].
  
 \bibitem{Ciuchini:2017mik}
  M.~Ciuchini, A.~M.~Coutinho, M.~Fedele, E.~Franco, A.~Paul, L.~Silvestrini and M.~Valli,
  Eur.\ Phys.\ J.\ C {\bf 77} (2017) no.10,  688
  [arXiv:1704.05447 [hep-ph]].
  
 \bibitem{Geng:2017svp}
  L.~S.~Geng, B.~Grinstein, S.~J\"ager, J.~Martin Camalich, X.~L.~Ren and R.~X.~Shi,
  Phys.\ Rev.\ D {\bf 96} (2017) no.9,  093006
  [arXiv:1704.05446 [hep-ph]].
  

 \bibitem{Borzumati:1999sp}
  F.~Borzumati, G.~R.~Farrar, N.~Polonsky and S.~D.~Thomas,
  Nucl.\ Phys.\ B {\bf 555}, 53 (1999)
  [hep-ph/9902443].


 \bibitem{Martin:1999hc}
  S.~P.~Martin,
  Phys.\ Rev.\ D {\bf 61}, 035004 (2000)
  [hep-ph/9907550].
 \bibitem{ArkaniHamed:1998wc}
  N.~Arkani-Hamed and R.~Rattazzi,
  Phys.\ Lett.\ B {\bf 454}, 290 (1999)
  [hep-th/9804068].
\bibitem{Un:2014afa}
 C.~S.~\"{Un},  \c{S}.~H.~Tany\i ld\i z\i,S.~Kerman and L.~Solmaz,
  Phys.\ Rev.\ D {\bf 91}, no. 10, 105033 (2015)
  [arXiv:1412.1440 [hep-ph]].
  \bibitem{Chattopadhyay:2017qvh}
  U.~Chattopadhyay, D.~Das and S.~Mukherjee,
  JHEP {\bf 1801}, 158 (2018)
  [arXiv:1710.10120 [hep-ph]].
  \bibitem{Chattopadhyay:2016ivr}
  U.~Chattopadhyay and A.~Dey,
  JHEP {\bf 1610}, 027 (2016)
  [arXiv:1604.06367 [hep-ph]].



 \bibitem{Alonso:2017bff}
  R.~Alonso, P.~Cox, C.~Han and T.~T.~Yanagida,
  Phys.\ Rev.\ D {\bf 96}, no. 7, 071701 (2017)
  [arXiv:1704.08158 [hep-ph]].
  \bibitem{Alonso:2017uky}
 R.~Alonso, P.~Cox, C.~Han and T.~T.~Yanagida,
 Phys.\ Lett.\ B {\bf 774}, 643 (2017)
 [arXiv:1705.03858 [hep-ph]].

\bibitem{Altmannshofer:2014cfa}
  W.~Altmannshofer, S.~Gori, M.~Pospelov and I.~Yavin,
  Phys.\ Rev.\ D {\bf 89}, 095033 (2014)
  [arXiv:1403.1269 [hep-ph]].
  \bibitem{Duan:2017qwj}
  G.~H.~Duan, X.~G.~He, L.~Wu and J.~M.~Yang,
  Eur.\ Phys.\ J.\ C {\bf 78}, no. 4, 323 (2018)
  [arXiv:1711.11563 [hep-ph]].
\bibitem{Tang:2017gkz}
  Y.~Tang and Y.~L.~Wu,
  Chin.\ Phys.\ C {\bf 42}, no. 3, 033104 (2018)
  [arXiv:1705.05643 [hep-ph]].

\bibitem{Yin:2018qcs}
  W.~Yin,
  arXiv:1808.00440 [hep-ph].



\bibitem{Hook:2010tw}
  A.~Hook, E.~Izaguirre and J.~G.~Wacker,
  Adv.\ High Energy Phys.\  {\bf 2011}, 859762 (2011)
  [arXiv:1006.0973 [hep-ph]].
\bibitem{pdg}
  C.~Patrignani {\it et al.} [Particle Data Group],
  Chin.\ Phys.\ C {\bf 40}, no. 10, 100001 (2016).
\bibitem{Chankowski:2006jk}
  P.~H.~Chankowski, S.~Pokorski and J.~Wagner,
  Eur.\ Phys.\ J.\ C {\bf 47}, 187 (2006)
  [hep-ph/0601097].


\bibitem{Arnan:2016cpy}
  P.~Arnan, L.~Hofer, F.~Mescia and A.~Crivellin,
  JHEP {\bf 1704} (2017) 043
  [arXiv:1608.07832 [hep-ph]].
  

\bibitem{Madgraph}
  J.~Alwall {\it et al.},
  JHEP {\bf 1407}, 079 (2014).

\bibitem{Aaboud:2017buh}
  M.~Aaboud {\it et al.} [ATLAS Collaboration],
  JHEP {\bf 1710}, 182 (2017)
  [arXiv:1707.02424 [hep-ex]].
  
\bibitem{Demir:2005ti}
  D.~A.~Demir, G.~L.~Kane and T.~T.~Wang,
  Phys.\ Rev.\ D {\bf 72}, 015012 (2005)
  [hep-ph/0503290].
\bibitem{Cox:2018qyi}
  P.~Cox, C.~Han and T.~T.~Yanagida,
  arXiv:1805.02802 [hep-ph].
  



\end{thebibliography}
\end{document}